\documentclass[prl,reprint,nofootinbib,showpacs,twocolumn]{revtex4-1}
\usepackage{amsmath, amssymb,bm}
\usepackage[T1]{fontenc}

\begin{document}
\title{The not-so-nonlinear nonlinearity of Einstein's equation}
\author{Abraham I. Harte}
\affiliation{
 Max-Planck-Institut f\"ur Gravitationsphysik, Albert-Einstein-Institut
 \\ Am M\"uhlenberg 1, 14476 Golm, Germany.}
\email{harte@aei.mpg.de}

\begin{abstract}
Many of the technical complications associated with the general theory of relativity ultimately stem from the nonlinearity of Einstein's equation. It is shown here that an appropriate choice of dynamical variables may be used to eliminate all such nonlinearities beyond a particular order: Both Landau-Lifshitz and tetrad formulations of Einstein's equation are obtained which involve only finite products of the unknowns and their derivatives. Considerable additional simplifications arise in physically-interesting cases where metrics becomes approximately Kerr or, e.g., plane waves, suggesting that the variables described here can be used to efficiently reformulate perturbation theory in a variety of contexts. In all cases, these variables are shown to have simple geometrical interpretations which directly relate the local causal structure associated with the metric of interest to the causal structure associated with a prescribed background. A new method to search for exact solutions is outlined as well.   
\end{abstract}


\maketitle

\vskip 1pc

Textbook discussions of general relativity typically emphasize its geometric simplicity. Einstein's equation, $R^{a}{}_{b} - \frac{1}{2} \delta^{a}_b R^{c}{}_{c} = 8 \pi T^{a}{}_{b}$, directly relates the Ricci curvature $R^{a}{}_{b}$ of spacetime to the stress-energy tensor $T^{a}{}_{b}$ of any matter fields which may be present. Although this relation is linear and algebraic, it can nevertheless be difficult to apply in practice. Part of the reason for this is that far more than $R^{a}{}_{b}$ is needed to fully characterize a particular spacetime. One might also be interested in the spacetime metric $g_{ab}$, its associated gradient $\nabla_a$, the Weyl curvature, etc. All of these structures may be computed from $g_{ab}$, suggesting that it can be useful to interpret Einstein's equation not as an algebraic constraint on the curvature, but rather as a differential equation for the metric. The precise form of this differential equation depends on, e.g., which variables are used to parametrize the metric: coordinate components, a tetrad, a $3+1$ decomposition, or something else. 

This Letter considers a particular split of $g_{ab}$ into a ``background'' piece $\hat{g}_{ab}$ and appropriate ``deviations.'' In terms of the natural variables associated with this decomposition, all high-order nonlinearities in Einstein's equation are shown to vanish. Furthermore, \textit{all} nonlinearity disappears in physically-interesting limits such as those described by the Schwarzschild or Kerr geometries, or by plane-symmetric gravitational waves. These variables are most directly of interest for simplifying analytic work on perturbation theory, but can also arise in connection with certain non-perturbative concepts.

The results described here are motivated by a theorem originally obtained by Llosa and Soler \cite{LlosaSoler}, following a conjecture due to Coll \cite{CollConj} which attempted to characterize the gauge-independent ``degrees of freedom'' present in a general $n$-dimensional metric (independently of any dynamical equations which may be imposed). The number of such degrees of freedom might be computed by first counting the number of independent entries in a symmetric $n \times n$ matrix, and then subtracting $n$ to account for gauge degrees of freedom. The resulting $\frac{1}{2} n(n-1)$ degrees of freedom is also the number of degrees of freedom associated with an arbitrary 2-form $f_{ab} = f_{[ab]}$ in $n$ dimensions, suggesting that metric degrees of freedom might be representable in terms of 2-forms. Indeed, Llosa and Soler showed that together with a conformal factor, a general analytic metric can, at least in finite regions, be written as a constant-curvature metric plus the square of an appropriate $f_{ab}$ \cite{LlosaSoler}. For the $n=4$ case of interest here, a particularly simple expression was shown to follow by expanding $f_{ab}$ in terms of its principal null directions \cite{LlosaCarot}. Generalizing slightly, a \textit{flat} background $\hat{g}_{ab}$ can be deformed into a generic $g_{ab}$ using only a conformal factor $\Omega$ and a pair of 1-forms $\ell_a$, $k_a$ which are both null with respect to $\hat{g}_{ab}$: 
\begin{equation}
    g_{ab} = \Omega^2 ( \hat{g}_{ab} + 2 \ell_{(a} k_{b)} ).
    \label{xKS}
\end{equation}
This decomposition cleanly splits the metric perturbation into parts which do and do not affect the causal structure of $g_{ab}$. The factorized perturbation $h_{ab} \equiv 2 \ell_{(a} k_{b)}$ describes five degrees of freedom which deform the light cones of $g_{ab}$ relative to those of $\hat{g}_{ab}$. Combining this with $\Omega$, which does not affect light cones, provides a total of six metric degrees of freedom. The remainder of this Letter explores various consequences of describing metric differences in terms of $\Omega$, $\ell_a$, and $k_a$.



One such consequence is that $h_{ab}$ must be proportional to a projection operator. Letting $h \equiv h^{a}{}_{a}$ while using $\hat{g}_{ab}$ to raise and lower indices both here and below,
\begin{equation}
    h_{a}{}^{c} h_{bc} = \frac{1}{2} h h_{ab}.
    \label{Proj}
\end{equation}
It follows that the inverse metric $g^{ab}$ ($\neq \hat{g}^{ac} \hat{g}^{bd} g_{cd}$) is
\begin{equation}
    g^{ab} = \Omega^{-2} \big[ \hat{g}^{ab} - \big( 1+\frac{1}{2} h \big)^{-1} h^{ab} \big],
    \label{gInv}
\end{equation}
which implies that $\ell_a$ and $k_a$ are null with respect to $g_{ab}$ as well as $\hat{g}_{ab}$. In each tangent space, the light cones of both metrics therefore coincide along the rays tangent to $\ell^a$ and $k^a$. If these preferred vectors are not proportional, $h$ is nonzero and can be viewed as measuring the local deformation of one light cone with respect to the other. The null vectors of $g_{ab}$ lie almost entirely inside those of $\hat{g}_{ab}$ wherever $-2 < h<0$, while the opposite is true wherever $h>0$. 

Another interpretation for $h$ is that it describes how volume measurements differ with respect to both metrics. Defining the ratio of the volume elements associated with $g_{ab}$ and $\hat{g}_{ab}$ as the proportionality factor between $\epsilon_{abcd}$ and $\hat{\epsilon}_{abcd}$,
\begin{align}
    \frac{ \sqrt{ - g } } { \sqrt{ - \hat{g} } } = \Omega^4 \big( 1 + \frac{1}{2} h \big).
    \label{Detg}
\end{align}
As the notation suggests, this reduces to a ratio of determinants in any coordinate system. If the conformal factor is ignored, \eqref{Detg} coincides exactly with the expression usually obtained in linearized perturbation theory. It is nevertheless exact for metrics with the form \eqref{xKS}.

Identities \eqref{Proj}-\eqref{Detg} greatly simplify Einstein's equation. One way to demonstrate this explicitly is to consider the Landau-Lifshitz formulation, which assumes a flat background and introduces the ``gothic metric'' 
\begin{equation}
    \mathfrak{g}^{ab} \equiv \left( \frac{\sqrt{-g}}{\sqrt{-\hat{g}}} \right) g^{ab} = \Omega^2 \big[ (1 + \frac{1}{2} h) \hat{g}^{ab} - h^{ab} \big].
    \label{gGoth}
\end{equation}
Applying only the first (defining) equality here, Einstein's equation can be written in terms of the background derivative $\hat{\nabla}_a$ as
\begin{align}
  \hat{\nabla}_c \hat{\nabla}_d ( \mathfrak{g}^{a[b} \mathfrak{g}^{c]d} ) = 8 \pi (g/\hat{g}) \left( g^{bc} T^{a}{}_{c} + t^{ab}_\mathrm{LL} \right),
  \label{LL}
\end{align}
where the Landau-Lifshitz tensor $t^{ab}_\mathrm{LL}$ satisfies
\begin{align}
    16 \pi (g/\hat{g}) t^{ab}_\mathrm{LL} \equiv g_{cd} ( g^{ef} \hat{\nabla}_e \mathfrak{g}^{ac} \hat{\nabla}_f \mathfrak{g}^{bd} - 2 g^{e(a} \hat{\nabla}_f \mathfrak{g}^{b)c} \hat{\nabla}_e \mathfrak{g}^{df} )
    \nonumber
    \\
    ~ + \frac{1}{8} ( 2 g_{cd} g_{ef} - g_{de} g_{cf} ) ( 2 g^{a p} g^{b q} - g^{ab} g^{pq}) \hat{\nabla}_p \mathfrak{g}^{cf} \hat{\nabla}_q \mathfrak{g}^{de} 
    \nonumber
    \\
    ~ + 2 \hat{\nabla}_c \mathfrak{g}^{a[b} \hat{\nabla}_d \mathfrak{g}^{c]d} + \frac{1}{2} g_{cd} g^{ab} \hat{\nabla}_e \mathfrak{g}^{cf} \hat{\nabla}_f \mathfrak{g}^{de} .
    \label{tLL}
\end{align}
Together with inertial coordinates and the harmonic gauge condition $\hat{\nabla}_b \mathfrak{g}^{ab} = 0$ [which is not generally compatible with \eqref{xKS}], these equations are most commonly used as the starting point for the post-Minkowski and post-Newtonian approximations to general relativity. In those contexts, \eqref{LL} is typically viewed as a differential equation for the perturbation $\mathfrak{h}_{ab} \equiv \mathfrak{g}^{ab} - \hat{g}^{ab}$ \cite{PoissonWill}. Formally expanding it in powers of $\mathfrak{h}_{ab}$ yields an infinite series, a result which is sometimes taken to imply that Einstein's equation involves nonpolynomial nonlinearities.

Such arguments are misleading. The potentially troublesome terms are those in $t_\mathrm{LL}^{ab}$ which involve $g_{ab} g^{cd}$. While such factors are indeed nonpolynomial in $\mathfrak{h}^{ab}$, they are rational. The relevant denominators involve only powers of $g/\hat{g}$, and may be eliminated by multiplying both sides of \eqref{LL} by a sufficiently large power of this factor. Related methods for eliminating nonpolynomial nonlinearities in Einstein's equation have been discussed in, e.g., \cite{Peres, Katanaev}. Unfortunately, they give rise to very complicated equations, and also provide no clear 
simplifications at lower orders in perturbation theory.

Considerably simpler results arise in connection with \eqref{xKS}. Using this decomposition, it follows immediately from \eqref{gInv}-\eqref{gGoth} that all non-matter terms in \eqref{LL} are rational in $h_{ab} = 2 \ell_{(a} k_{b)}$ and $\ln \Omega$. Converting them into polynomials requires only multiplication by $\Omega^{-4} (1+\frac{1}{2} h)$. Indeed, $\Omega^{4} (1+ \frac{1}{2} h)^{3} t^{ab}_\mathrm{LL}$ includes only terms which involve between two and five powers of the unknowns and their derivatives. The vacuum Einstein equation as a whole involves terms with between one and five powers of the unknowns. Letting $C_{cab} \equiv \hat{\nabla}_{(a} h_{b)c} - \frac{1}{2} \hat{\nabla}_c h_{ab}$ and $T_{ab}^\mathrm{TR} \equiv (g_{ac} \delta^d_b - \frac{1}{2} g_{ab} \delta_c^d)T^{c}{}_{d}$, the full (not necessarily vacuum) Einstein equation may be explicitly written as
\vskip -.5 cm    
\begin{widetext}
\begin{align}
    8\pi(1+\frac{1}{2} h)^2 T_{ab}^{\mathrm{TR}}  = (1 + \frac{1}{2} h)  \Big\{ (\Omega^{-2} \mathfrak{g}^{cd}) \Big[ \hat{\nabla}_c \hat{\nabla}_{(a} h_{b)d} - \frac{1}{2} \hat{\nabla}_c \hat{\nabla}_d h_{ab} - (\hat{g}_{ab} + h_{ab}) (\hat{\nabla}_c \hat{\nabla}_d \ln \Omega + 2 \hat{\nabla}_c \ln \Omega \hat{\nabla}_d \ln \Omega) 
    \nonumber
    \\
     + 2 C_{cab} \hat{\nabla}_d \ln\Omega \Big] - (\hat{g}^{cf} \hat{g}^{de} - \frac{1}{2} \hat{g}^{cd} \hat{g}^{ef} ) C_{cab} \hat{\nabla}_d h_{ef} - \frac{1}{2} \hat{\nabla}_a \hat{\nabla}_b h + (\hat{g}_{ab} + h_{ab}) C^{cd}{}_{d} \hat{\nabla}_c \ln \Omega - 2 (1 + \frac{1}{2} h) 
    \nonumber
    \\
    ~ ( \hat{\nabla}_a \hat{\nabla}_b \ln \Omega - \hat{\nabla}_a \ln \Omega \hat{\nabla}_b \ln \Omega)   \Big\}  + (\Omega^{-2} \mathfrak{g}^{cd}) (\Omega^{-2} \mathfrak{g}^{ef}) \big( \hat{\nabla}_c h_{ae} \hat{\nabla}_{[d} h_{f]b} - \frac{1}{4} \hat{\nabla}_a h_{ce} \hat{\nabla}_b h_{df} \big) + \frac{1}{4} \hat{\nabla}_a h \hat{\nabla}_b h  .
    \label{Einstein}
\end{align}
\end{widetext}
Noting that the right-hand side of this expression is equal to $(1 + \frac{1}{2}h)^2 g_{ac} R^{c}{}_{b}$, its trace may be used to show that the Einstein-Hilbert action $(\sqrt{-g}/\sqrt{-\hat{g}} )R^{a}{}_{a}$ has the form $\Omega^2 (1+ \frac{1}{2} h)^{-2} \times (\mbox{1st through 5th order terms})$.


 
In general, the ability to remove nonpolynomial nonlinearities in \eqref{LL} may allow the formulation of perturbative schemes with simpler properties at high orders. Perhaps more interestingly, however, the metric decomposition adopted here can provide simplifications even at low orders in perturbation theory. There is a sense in which it can, e.g.,  ``resum'' perturbative series in useful ways.

Even the transformation of a known approximate metric into the form \eqref{xKS} can result, in a natural way, in a new metric which is much more accurate than its ``seed.'' The Newtonian approximation to general relativity provides an example of this. Given a Newtonian potential $\Phi_\mathrm{N}$, it is standard to consider the metric $g^{\mathrm{N}}_{ab} = (1 - 2 \Phi_\mathrm{N} ) \hat{g}_{ab} - 4 \Phi_\mathrm{N} \nabla_a t \nabla_b t$, where $\hat{g}_{ab}$ is a flat background and $t$ is an inertial time coordinate normalized such that $\hat{g}^{ab} \nabla_a t \nabla_b t = -1$. This is not in the form \eqref{xKS}, so consider instead those metrics $g_{ab} = g_{ab}^\mathrm{N} +  \mathcal{L}_\xi \hat{g}_{ab}$ which can be obtained from $g_{ab}^\mathrm{N}$ via first-order gauge transformations. Cases where $\Phi_\mathrm{N}$ describes a static, spherically-symmetric star with finite radius are particularly straightforward. If such a star is not too compact, explicit gauge vectors $\xi^a$ may be found such that i) $g_{ab}$ is globally in the form \eqref{xKS} with $\ell_a \propto k_a$, ii) outside of the star, $g_{ab}$ is \textit{exactly} Schwarzschild, iii) the relativistic and Newtonian masses are exactly equal, iv) the Newtonian radius of the star is exactly equal to the relativistic areal radius, and v) the Newtonian mass density $\rho_\mathrm{N}$ is related to the relativistic rest energy density $\rho$ and the principal pressures $p_i$ via $\rho_\mathrm{N} = ( \sqrt{-g} / \sqrt{-\hat{g}} ) ( \rho + \sum_i p_i)$. The gauge vectors which accomplish this depend only \textit{linearly} on $\Phi_\mathrm{N}$. The metric decomposition \eqref{xKS} therefore suggests a simple map between spherical solutions in Newtonian gravity and spherical solutions in full general relativity. Somewhat weaker results are expected to persist even in certain non-spherical systems.

These and other simplifications associated with \eqref{xKS} are related to the special properties of limits where $h \rightarrow 0$. If $h_{ab}$ does not vanish in such a limit, $\ell_a$ and $k_a$ must become parallel. After an appropriate rescaling, it then follows that $g_{ab} \rightarrow \Omega^2 ( \hat{g}_{ab} \pm 2 \ell_a \ell_b)$. At least when $\Omega = 1$, metrics with this form are described as being of Kerr-Schild type\footnote{Cases where $\Omega$ is nontrivial while $h_{ab} = 0$ are also interesting as, e.g., limiting geometries on cosmological scales.}. As already noted in \cite{LlosaCarot}, the decomposition \eqref{xKS} may be interpreted as a generalization of the Kerr-Schild ansatz. Unlike either the original Kerr-Schild ansatz or its other generalizations \cite{Taub, Bonanos, EttKastor}, however, the form discussed here is known to encompass very general geometries. The classical Kerr-Schild metrics nevertheless represent an important special case. All vacuum examples are known \cite{ExactSolns}, and there are even senses in which Einstein's equation becomes linear within this class \cite{Xanthopoulos1, Xanthopoulos2, Xanthopoulos3, Gursey, Gergely, Taub, ExactSolns}. 

The Kerr-Schild metrics are also important physically. At least in vacuum, many of the most important exact solutions which are known are members of this class. Spherically-symmetric vacuum solutions---which must be Schwarzschild---are Kerr-Schild, for example. Indeed, all (not necessarily vacuum) static, spherically-symmetric metrics are at least conformal to Kerr-Schild metrics \cite{ConfKS}. The rotating Kerr black holes are Kerr-Schild as well, and arise as the endpoints of, e.g., black hole collisions and certain types of gravitational collapse. Even before two black holes collide, the geometry is very nearly Kerr-Schild in the vicinity of each black hole (an observation which has been exploited to construct initial data for numerical simulations \cite{Bishop, Sarbach}). Similar statements can also apply far away from generic physical systems. Moving in spacelike directions, isolated asymptotically-flat spacetimes look increasingly like Schwarzschild at large distances, indicating that $h$ tends to zero more rapidly in such directions than $h_{ab}$ as a whole. Also in the Kerr-Schild class are gravitational plane waves, suggesting that $h$ may decay more rapidly than $h_{ab}$ even when moving away from isolated systems along null directions.

These observations suggest that the metric decomposition considered here can be used as a new analytic tool with which to learn about black hole binaries and similar systems. Current understanding of such problems draws from a combination of analytic approximations, full numerical simulations of Einstein's equation, and phenomenological modeling \cite{AlexBBH}. From the perspective of \eqref{xKS}, $h_{ab}$ can always be split into a ``Kerr-Schild component'' $K_{ab} \equiv 2 \ell_a \ell_b$ plus a correction $X_{ab} \equiv 2 \ell_{(a} \zeta_{b)}$, where scalings are chosen such that $\ell_{[a} \zeta_{b]} \neq 0$ unless $X_{ab} = 0$. Except perhaps during the final collision between two black holes, $X_{ab}$ is expected to remain small everywhere, the majority of the strong-field behavior being accounted for by $K_{ab}$ alone. Noting that the strong-field behavior is isolated in this way is useful because of the different ways in which $K_{ab}$ and $X_{ab}$ appear in Einstein's equation. Even if $K_{ab}$ is large and not put in ``by hand'' (as effectively occurs in typical formulations of black hole perturbation theory), good approximations to it can arise dynamically even at low orders in a standard nonlinearity expansion. As a simple example, such expansions can admit the \textit{exact} Kerr solution as a first-order perturbation to flat spacetime; the strong-field structure of the metric appears all at once, rather than bit by bit at each order. Moreover, additional perturbations described by nonzero $X_{ab}$ or corrections to $K_{ab}$ couple only in relatively simple ways to the lowest-order $K_{ab}$. Nonlinearities involving $X_{ab}$ alone can be more complicated, although the expected smallness of this field is likely to make such effects ignorable in many cases.

The sense in which Einstein's equation treats $K_{ab}$ and $X_{ab}$ differently may be understood more precisely by first raising one of the indices in \eqref{Einstein} with $g^{ac}$, and then undoing the trace-reversal employed there. This results in an equation for $\Omega^2 (1+\frac{1}{2} h)^3 T^{c}{}_{b}$ which involves terms containing between one and five powers of $h_{ab}$ and $\ln \Omega$. No term in that equation contains more than four powers of $h_{ab}$. Substituting $h_{ab} = K_{ab} + X_{ab}$ shows, however, that only two powers of $K_{ab}$ can appear. The Kerr-Schild component of an arbitrary perturbation therefore couples relatively little both to itself and to $X_{ab}$. Further simplifications arise if $\ell^b \hat{\nabla}_b \ell^a \propto \ell^a$, which is the case for any purely Kerr-Schild solution in vacuum: All terms in Einstein's equation which are quadratic in $K_{ab}$ and independent of $X_{ab}$ then vanish identically. This generalizes one of the senses in which Einstein's equation is known \cite{Gursey} to become linear for Kerr-Schild metrics.

A detailed perturbative scheme which takes advantage of these results is not pursued here. We instead consider the problem that the ``relaxed'' Einstein equation \eqref{Einstein} is difficult to solve as-is. Viewed as an equation for $h_{ab}$ and $\ln \Omega$, it is not hyperbolic. Solutions also exist which do not satisfy the ``gauge condition'' $h_{ab} = 2 \ell_{(a} k_{b)}$, and therefore fail to be physically relevant. A more systematic approach would involve expanding everything explicitly in terms of $\ell_a$ and $k_a$. This is most naturally accomplished not by working directly from \eqref{Einstein}, but rather by considering a tetrad formulation of general relativity. This replaces the metric as the basic variable with a set of four linearly-independent vector fields (from which $g_{ab}$ can easily be reconstructed if desired). The specific approach adopted here is originally due to Geroch, Held, and Penrose (GHP) \cite{ExactSolns, GHP, PenroseRindler}, and is a refinement of the better-known Newman-Penrose formalism \cite{NP} adapted to systems where there exist two preferred null directions. It provides a viewpoint which is complementary to the tensorial one discussed above.

As a brief review, the GHP formalism replaces $g_{ab}$ by a complex null tetrad $(\ell^a, n^a, m^a, \bar{m}^a)$ normalized such that all inner products vanish except for $g_{ab} m^a \bar{m}^b = -g_{ab} \ell^a n^b = 1$ [so $g^{ab} = 2 (m^{(a} \bar{m}^{b)} -\ell^{(a} n^{b)})$]. Given any complex $\lambda \neq 0$, it is then interesting to consider those rotations and boosts
\begin{equation}
    \ell^a \rightarrow \lambda \bar{\lambda} \ell^a, \quad n^a \rightarrow (\lambda \bar{\lambda})^{-1} n^a, \quad m^a \rightarrow \lambda \bar{\lambda}^{-1} m^a
    \label{Boost}
\end{equation}
which preserve the directions of $\ell^a$ and $n^a$, but not necessarily those of $m^a$ and $\bar{m}^a$. Scalars $\eta$ which transform like $\eta \rightarrow \lambda^{b+s} \bar{\lambda}^{b-s} \eta$ are said to have boost weight $b$ and spin weight $s$. All first derivatives of the tetrad elements which have well-defined spin and boost weights are collected into eight complex ``spin coefficients'' $\kappa$, $\sigma$, $\rho$, $\tau$, $\kappa'$, $\sigma'$, $\rho'$, and $\tau'$. Primes denote a generic operation which effects the replacements $\ell^a \leftrightarrow n^a$ and $m^a \leftrightarrow \bar{m}^a$ in all definitions, so, e.g., $\kappa = -g_{bc} \ell^a m^b \nabla_a \ell^c$ and $\kappa' = -g_{bc} n^a \bar{m}^b \nabla_a n^c$. A complete set of derivative operators $\eth$ (eth), $\text{\th}$ (thorn), $\eth'$, and $\text{\th}'$ is also defined. These too have well-defined spin and boost weights. Together, the spin coefficients and the derivative operators completely determine the spacetime curvature \cite{GHP}.

Einstein's equation is imposed in this formalism by restricting the Ricci curvature, which in turn implies restrictions on the spin coefficients and the derivative operators (see \cite{Edgar80, Edgar92, GHPSeries, HeldGHP}). The result is essentially a first-order formulation of Einstein's equation. Components of the stress-energy tensor are expressed as first derivatives of the spin coefficients, and the spin coefficients are first derivatives of the tetrad. The former set of equations is relatively simple, involving only quadratic nonlinearities in the spin coefficients. Obtaining a tetrad from a known set of spin coefficients can be complicated, however. Some of these complications are bypassed if tetrads associated with $g_{ab}$ are expressed as deformations of known tetrads associated with a background metric. 

The decomposition \eqref{xKS} provides null vector fields which express such deformations in a particularly simple way. In this context, it is simplest to exclude purely Kerr-Schild or conformally Kerr-Schild metrics at the outset\footnote{This has the unfortunate side-effect that limits to nontrivial Kerr-Schild spacetimes become singular. Such cases might be better understood by instead adopting a version of the GHP formalism which does not require that the tetrad be normalized \cite{PenroseRindler}. An approach where only one tetrad direction is fixed \cite{Vickers} might also be useful.}, so $h \neq 0$. Letting $\chi \equiv \Omega (1+ \frac{1}{2} h)^{1/2}$, the directions of $\ell^a$ and $k^a$ can then be identified with the first two elements of a \textit{background} GHP tetrad via $\hat{\ell}^a = \chi \ell^a$ and $\hat{n}^a = - 2 (\chi h)^{-1} k^a$. Supplementing these vectors with any appropriate $\hat{m}^a$, an admissible GHP tetrad for $g_{ab}$ can be obtained by simple rescalings:
\begin{equation}
    \ell^a = \chi^{-1} \hat{\ell}^a, \quad n^a = \chi^{-1} \hat{n}^a, \quad m^a = \Omega^{-1} \hat{m}^a.
    \label{dTetrad}
\end{equation}
Elements of physical null tetrads which are adapted to \eqref{xKS} are therefore proportional to elements of appropriate \textit{flat} tetrads. The scalings here involve only the two scalar fields $\Omega$ and $\chi$, or equivalently $\ln \Omega$ and $h$.

Direct computation using \eqref{xKS}-\eqref{gInv} and the definitions found in \cite{GHP} now shows that
\begin{subequations}
\label{Spin}
\begin{align}
    \kappa &= \Omega^{-1} \hat{\kappa} , \qquad \sigma = \chi^{-1} \hat{\sigma},
    \label{Spin1}
    \\
    \rho &= \chi^{-1} \left[ \hat{\rho} + \frac{1}{4} h ( \hat{\rho} - \hat{\bar{\rho}} ) - \hat{ \text{\th} } \ln \Omega \right],
    \label{Spin2}
    \\
    \tau &= \Omega^{-1} \left[ \hat{\tau} - \frac{1}{4} h (\Omega/\chi)^2 ( \hat{\tau} -\hat{\bar{\tau}}' ) - \hat{\eth} \ln \chi \right].
    \label{Spin3}
\end{align}
\end{subequations}
The background and perturbed spin coefficients are therefore related by a linear transformation depending on $\Omega$ and $\chi$, plus simple inhomogeneous terms depending linearly on derivatives of these two scalars. Similar expressions relate the background and perturbed derivative operators. When acting on any scalar with boost weight $b$ and spin weight $s$,
\begin{subequations}
\label{Derivatives}
\begin{align}
    \text{\th} &= \chi^{-1} \big[ \chi^b \hat{ \text{\th} } \chi^{-b} 
    - \frac{1}{4} s h ( \hat{\rho} - \hat{\bar{\rho}} ) \big] ,
    \label{thorn}
    \\
    \eth &= \Omega^{-1} \big[ \Omega^s \hat{\eth} \Omega^{-s} + \frac{1}{4} b h (\Omega/\chi)^2 ( \hat{\tau} - \hat{\bar{\tau}}' )  \big].
    \label{eth}
\end{align}
\end{subequations}
All of these equations have primed counterparts in which, e.g., $( b, s ) \rightarrow ( -b , -s )$. One consequence of \eqref{Spin} is that $\kappa$, $\kappa'$, $\sigma$, $\sigma'$, $\rho - \bar{\rho}$, $\rho' - \bar{\rho}'$, and $\tau - \bar{\tau}'$ are merely rescalings of their background counterparts. Various relations between the optical scalars associated with the background and perturbed $\ell^a$ and $n^a$ follow immediately. 

Using \eqref{Spin} and \eqref{Derivatives}, Einstein's equation can now be solved via the curvature and commutator relations obtained in \cite{GHP}. The unknowns in this context are $h$, $\ln \Omega$, and a flat tetrad. Any two flat tetrads can be related by a Lorentz transformation, so the unknown tetrad here may be parametrized via a Lorentz transformation which acts on a convenient fiducial tetrad. Alternatively, a particular flat tetrad, or a class of them, could be fixed beforehand. All differential relations between the spin coefficients and the perturbed tetrad are then taken into account automatically by \eqref{dTetrad} and \eqref{Spin}. Only those equations which relate the spin coefficients to the stress-energy tensor must be solved. Indeed, it is straightforward to compute, e.g., $T^{a}{}_{b}$ as a function of $\ln \Omega$, $h$, and any free parameters in the class of chosen tetrads. Such computations are particularly well-suited to finding exact solutions.

Suppose, for example, that $g_{ab}$ is to solve the vacuum Einstein equation, and that $\ell^a$ is assumed to be a repeated principal null direction. It then follows from the Goldberg-Sachs theorem \cite{ExactSolns, GoldbergSachs} and \eqref{Spin1} that $\hat{\ell}^a$ must be null, geodesic, and shear-free with respect to $\hat{g}_{ab}$. All such vector fields are known via Kerr's theorem \cite{KerrThm1, KerrThm2}, so ``only'' those particular tetrads in this class must be found which, together with appropriate $h$ and $\ln \Omega$, imply that $R^{a}{}_{b} = 0$. In general, however, neither $\ell^a$ nor $n^a$ must be principal null vectors at all, let alone repeated ones. 

\begin{acknowledgments}

\vskip .1 cm
\noindent
\textit{Acknowledgements.} The author thanks Stanislav Babak and Yi-Zen Chu for comments, and also Thomas Linz and Alessandra Buonanno for useful suggestions. Some calculations were performed using the \texttt{xAct} tensor analysis package \cite{xAct} for \texttt{Mathematica}.

\end{acknowledgments}


\begin{thebibliography}{1}
\bibitem{LlosaSoler} J. Llosa and D. Soler, Class. Quantum Grav. \textbf{22}, 893 (2005)
\bibitem{CollConj} B. Coll in \textit{Gravitation and Relativity in General}, edited by J. Martin, E. Ruiz, F. Atrio, and A. Molina (World Scientific, Singapore, 1999)
\bibitem{LlosaCarot} J. Llosa and J. Carot, Class. Quantum Grav. \textbf{26}, 055013 (2009)
\bibitem{PoissonWill} E. Poisson and C. M. Will, \textit{Gravity: Newtonian, Post-Newtonian, Relativistic} (Cambridge University Press, Cambridge, 2014)
\bibitem{Peres} A. Peres, Nuovo Cimento \textbf{28}, 865 (1963)
\bibitem{Katanaev} M. O. Katanaev, Gen. Rel. Grav. \textbf{38}, 1233 (2006)
\bibitem{Taub} A. H. Taub, Ann. Phys. (NY) \textbf{134}, 326 (1981)
\bibitem{Bonanos} S. Bonanos, Class. Quantum Grav. \textbf{9}, 697 (1992)
\bibitem{EttKastor} B. Ett and D. Kastor, Class. Quantum Grav. \textbf{27}, 185024 (2010)
\bibitem{ExactSolns} H. Stephani, D. Kramer, M. MacCallum, C. Hoenselaers, and E. Herlt, \textit{Exact Solutions of Einstein's Field Equations} 2nd ed. (Cambridge University Press, Cambridge, 2003)
\bibitem{Gergely} L. \'{A}. Gergely, Class. Quantum Grav. \textbf{19}, 2515 (2002)
\bibitem{Gursey} M. G\"{u}rses and F. G\"{u}rsey, J. Math. Phys. \textbf{16}, 2385 (1975)
\bibitem{Xanthopoulos1} B. C. Xanthopoulos, J. Math. Phys. \textbf{19}, 1607 (1978)
\bibitem{Xanthopoulos2} B. C. Xanthopoulos, Class. Quantum Grav. \textbf{3}, 157 (1986)
\bibitem{Xanthopoulos3} K. E. Mastronikola and B. C. Xanthopoulos, Class. Quantum Grav. \textbf{6}, 1613 (1989)
\bibitem{ConfKS} N. V. Mitskievich and J. Horsk\'{y}, Class. Quantum Grav. \textbf{13}, 2603 (1996)
\bibitem{Bishop} N. T. Bishop, R. Isaacson, M. Maharaj, and J. Winicour, Phys. Rev. D \textbf{57}, 6113 (1998)
\bibitem{Sarbach} O. Sarbach, M. Tiglio, and J. Pullin, Phys. Rev. D \textbf{65}, 064026 (2002)
\bibitem{AlexBBH} A. Le Tiec, Int. J. Mod. Phys. D \textbf{23}, 1430022 (2014)
\bibitem{GHP} R. Geroch, A. Held, and R. Penrose, J. Math. Phys. \textbf{14}, 874 (1973)
\bibitem{PenroseRindler} R. Penrose and W. Rindler, \textit{Spinors and Spacetime} (Cambridge University Press, Cambridge, 1984)
\bibitem{NP} E. Newman and R. Penrose, J. Math. Phys. \textbf{3}, 566 (1962)
\bibitem{Edgar80} S. B. Edgar, Gen. Rel. Grav. \textbf{12}, 347 (1980)
\bibitem{Edgar92} S. B. Edgar, Gen. Rel. Grav. \textbf{24}, 1267 (1992)
\bibitem{GHPSeries} G. Ludwig and S. B. Edgar, Gen. Rel. Grav. \textbf{28}, 707 (1996)
\bibitem{HeldGHP} A. Held, Comm. Math. Phys. \textbf{37}, 311 (1974)
\bibitem{Vickers} M. P. Machado Ramos and J. A. G. Vickers, Class. Quantum Grav. \textbf{13}, 1579 (1996)
\bibitem{GoldbergSachs} J. N. Goldberg and R. K. Sachs, Acta Phys. Polon. Suppl. \textbf{22}, 13 (1962)
\bibitem{KerrThm1} R. Penrose, J. Math. Phys. \textbf{8}, 345 (1967)
\bibitem{KerrThm2} D. Cox and E. J. Flaherty Jr., Comm. Math. Phys. \textbf{47}, 75 (1976)
\bibitem{xAct} J.~M. Mart\'{i}n-Garc\'{i}a, \texttt{http://xact.es}
\end{thebibliography}
\end{document}